\documentstyle[twocolumn,aps]{revtex}
\begin{document}
\draft
\title{\bf {Mesoscopic superconducting disks}}
\author{P. Singha Deo\cite{eml}, F. M.  Peeters \cite{fmp}, and 
V. A. Schweigert\cite{sc}} 
\address{Department of Physics, University of Antwerp (UIA),
B-2610 Antwerpen, Belgium.}
\maketitle
\begin{abstract}
Using the non-linear Ginzburg-Landau (GL) eqs. type I
superconducting disks of finite radius ($R$) and thickness ($d$)
are studied in a perpendicular magnetic field. Depending on $R$
and $d$, first or second order phase transitions are found for
the normal to superconducting state. For sufficiently large $R$
several transitions in the superconducting phase are found
corresponding to different angular momentum giant vortex states.
In increasing magnetic field the superconductor is in its ground
state, while in field down sweep it is possible to drive the
system into metastable states. We also present a quantitative
analysis of the relation between the detector output and the
sample magnetization. The latter, and the incorporation of the
finite thickness of the disks, are essential in order to obtain
quantitative agreement with experiment.
\end{abstract}
\narrowtext

\section{Introduction}

Superconductivity is one of the most well studied phenomena in
condensed matter physics. Superconductivity in bulk samples,
thin films, cylinders and hollow cylinders have been extensively
studied and can even be found in text books \cite{boo}. With the
advent of nano-fabrication technologies, there is revived
interest in superconductivity in mesoscopic samples [2-10] and
to investigate the effect of the geometry and the size of the
sample in the superconducting state. Mesoscopic samples are
defined as samples whose size is comparable to the coherence
length. Moschalkov {\it et al} \cite{mos} measured the
superconducting-normal phase diagram of a mesoscopic
superconducting square and rectangular ring. They used
resistivity measurements which does not allow for probing deep
inside the superconducting state. Furthermore, the presence of
conduction probes may influence the superconducting state of the
studied sample.  In order to circumvent these limitations one
needs a contactless experiment which is provided by a
magnetization measurement.  Buisson {\it et al} \cite{bui}
performed magnetization measurements on an ensemble of $In$
disks with large separation between them in order to make the
dipolar interaction between the disks negligible. Recently Geim
{\it et al} \cite{gei} used sub-micron Hall probes to detect the
magnetization of {\it single} superconducting $Al$ disks with
size down to 0.1 $\mu m$.  These experimental works are the main
motivation behind our present and previous theoretical work.

Already a substantial amount of theoretical work has been done
on mesoscopic samples using the Ginzburg-Landau (GL) theory.
Some of these works neglect the non-linear term in the first GL
eqn \cite{buz,zw,bui}. But all these works neglect the
de-magnetization effect which arises due to the finite thickness
of the disk \cite{mae,lop,fom,pal,akk,mos1}.  The latter is very
important in case there is a substantial Meissner effect and
is the reason for the discrepancies between the experimental
observations and the existing theories.  For example, Buisson
{\it et al} \cite{bui} observed an oscillatory behavior (or
jumps) in the magnetization which occurs above a certain applied
magnetic field. Linearized Ginzburg-Landau (LGL) theory could
explain the origin of the oscillatory behavior, but is not able
to explain: $i$) the magnetic field above which the oscillations
occur, $ii$) its amplitude, and $iii$) the periodicity of these
oscillations.  Hence they questioned the validity of the GL
theory in dealing with the magnetization of mesoscopic samples.
Such oscillatory behavior has also been obtained in the single
disk measurement of Geim {\it et al}. \cite{gei}, provided the
disk is large enough to accommodate many flux quanta. Ref.
\cite{gei} showed that the very nature of the normal to
superconductor transition can change from a first order phase
transition to a second order as the dimensions of the disk is
varied.  Notice that bulk $Al$ has $\xi(0)=1.6 \mu m$, 
$\lambda(0)$=.016$\mu m$ and consequently the GL parameter is
$\kappa$=0.01 which tells us that $Al$ is a type I sample and
consequently the superconducting to normal transition is first
order in that case.

We showed recently that the GL theory can explain the magnetic
field at which the oscillations start, the amplitude of the
oscillations and the periodicity of the oscillations, provided
the non-linear term and the de-magnetization effect are included
in the theoretical description \cite{deo1,deo2}. 

The GL theory does not have a firm mathematical derivation
except in a narrow range of magnetic field close to the phase
boundary of type II bulk materials \cite{gor}.  However because
of its simplicity, it enables us to make quantitative
calculations for most of the experimentally observed quantities,
which makes it a popular theory. From experience it is known
that it gives sensible results even beyond this limited regime.
For example it was found to work for type I thin films
\cite{dol}. The reason is that thin films are governed by  an
effective penetration length $\lambda_e$= $\lambda^2/d$ which
results in an effective GL penetration length larger than
$1/\sqrt 2$. Surface superconductivity or the giant vortex state
could be quantitatively understood from the GL theory for
samples with boundaries \cite{deg}.  The phase boundary of
mesoscopic type I samples could be quantitatively explained from
the GL theory \cite{mos}. Refs. \cite{deo1,deo2,schweigert2} applied the GL
theory to study the magnetization of mesoscopic disks deep
inside the phase boundary and there was remarkable quantitative
agreement with the experimental observations.

In the present work the GL approach to disks is reviewed and the
non-linear Meissner effect is analyzed in detail.  We show that
the detector size greatly influences the measured magnetization.
In smaller disks it only influences the magnitude whereas in
larger disks it can influence both the magnitude as well as the
line shape. Thus we find that one has to be careful to translate
the measured Hall voltage directly into the sample
magnetization.  Also the hysteresis behavior of the
magnetization will be analyzed in detail.

\section{Theoretical treatment}

We consider superconducting disks with radius $R$ and thickness
$d$ immersed in an insulating medium. As a first approximation,
neglecting the non-local effects, we solve the system of two
coupled GL eqns.
\begin{equation}
\label{eq1}
\frac{1}{2m}\left(-i\hbar\vec \nabla -\frac{2e\vec A}{c}\right)^2\Psi=
-\alpha\Psi-\beta\Psi|\Psi|^2,
\end{equation}
\begin{equation}
\label{eq2}
\vec \nabla \times \vec \nabla\times \vec A=\frac{4\pi}{c}\vec j,
\end{equation}
where the density of superconducting current $\vec j$ is given by
\begin{equation}
\label{eq3}
\vec j=\frac{e\hbar}{im}
\left(\Psi^*\vec \nabla \Psi-\Psi \vec\nabla\Psi^*\right)
-\frac{4e^2}{mc}|\Psi|^2\vec A.
\end{equation}
Here $\alpha$ is a temperature dependent
parameter: $\alpha=\alpha(0)(T/T_c-1)$ while
$\beta$ is regular at $T_c$, where $T_c$ is the critical
temperature.  On the disk surface we require that the normal
component of the current density is zero
\begin{equation}
\label{eq4}
\left(-i\hbar\vec\nabla-\frac{2e\vec A}{c}\right)_n \Psi =0,
\end{equation}
The boundary condition for the vector potential is such that far
away from the superconducting disk the field equals the applied
field $\vec H=(0,0,H_0)$, i.e., $\vec A|_{\vec \rho\rightarrow
\infty}=\vec e_{\phi}H_0\rho /2$. Here $\vec e_{\phi}$ denotes
the azimuthal direction and $\rho$ is the radial distance from
the disk center.

Using the London gauge $div \vec A=0$, we rewrite the system of eqns.
(1-3) into the following form
\begin{equation}
\label{eqn1}
\left(-i\vec \nabla -\vec A\right)^2\Psi=
\Psi (1-|\Psi|^2),
\end{equation}
\begin{equation}
\label{eqn2}
-\kappa^2\triangle \vec A=
\frac{1}{2i}
\left(\Psi^*\vec \nabla \Psi-\Psi \vec\nabla\Psi^*\right)-|\Psi
|^2\vec A,
\end{equation}
where distance is measured in units of the coherence length
$\xi=\hbar/\sqrt{-2m\alpha}$, the order parameter in $\psi
_0=\sqrt{-\alpha/\beta}$, the vector potential in
$c\hbar/2e\xi$, $\kappa =\lambda /\xi$ is the GL parameter, and
$\lambda=c\sqrt{m/\pi}/4e\psi _0$ is the bulk penetration
length. We measure the magnetic field in $H_{c2}=c\hbar/2e\xi
^2=\kappa \sqrt{2}H_c$, where $H_c=\sqrt{-4\pi \alpha/\beta}$ is
the bulk critical field.  Notice that the parameters $\alpha$
and $\beta$ are scaled out of the GL equations and consequently
they only determine the energy and length scales.

The difference of the Gibbs free energy $G$ between the
superconducting and the normal state, measured  in
$H_c^2V/8\pi$, can be expressed through the integral
\begin{equation}
\label{free}
G=\int \left( 2(\vec A-\vec A_0).\vec j-|\Psi|^4\right)d\vec r/V,
\end{equation} 
over the disk volume $V=\pi R^2d$, where $\vec A_0=\vec
e_{\phi}H_0\rho/2$ is the external vector potential, and $\vec
j=(\Psi^*\vec \nabla \Psi-\Psi \vec\nabla\Psi^*)/2i -|\Psi
|^2\vec A$ is the dimensionless superconducting current. The
magnetization is defined as
\begin{equation}
M=\int_S(H_z-H_0)d\vec r/4 \pi,
\end{equation}
which is a direct measure of the expelled magnetic field from
the surface area $S$ of the disk.

We used the Gauss-Seidel method to solve Eqn. 1 and the fast
Fourier transform to solve Eqn 2. The order parameter of the
previous magnetic field is taken as the initial order parameter
for a particular magnetic field. A large number of iterations
are then made to arrive at a self consistent solution. In
previous work we found that the assumption that the order
parameter is uniform in the z-direction is a very good
approximation for the thin disks considered here
\cite{sch,deo1,deo2}.  Consequently, as far as the order
parameter is concerned the disk is reduced to a 2D disk. But the
3D nature of the field distribution is completely retained.
Hence Eqs. 1 and 2 becomes
\begin{equation}
\label{eq1}
\left(-i\vec \nabla_{2D} -\vec A\right)^2\Psi=
\Psi (1-|\Psi|^2),
\end{equation}
\begin{equation}
\label{eq2}
-\triangle_{3D} \vec A=\frac{d}{\kappa^2}\delta(z) \vec j_{2D},
\end{equation}
\begin{equation}
\label{eq3}
\vec j_{2D}=\frac{1}{2i}
\left(\Psi^*\vec \nabla_{2D} \Psi-\Psi \vec\nabla_{2D}\Psi^*\right)
-|\Psi |^2\vec A,
\end{equation}
with the boundary condition $(-i\vec \nabla_{2D}-\vec
A)_n\Psi|_{r=R}=0.$

It is known that for superconducting cylinders,  the sample can
only exhibit the Abrikosov vortex state when the GL parameter
$\kappa > 1/\sqrt 2$. This is no longer so for thin disks,
because they have an {\it effective} $\kappa$ which is much
different from that of cylinders. Outside the Abrikosov vortex
state region, both the Meissner state and the giant vortex state
have azimuthal symmetry. Hence for a disk that does not exhibit
the multi-vortex state \cite{schweigert2}, we can assume axially symmetric
solutions with a general angular momentum: $\Psi(\vec
\rho)=F(r)exp(iL\phi)$.  This will reduce the dimensions of Eqs.
1 and 2 and thus it greatly improves the computation time and
the accuracy.  For simplicity we will refer to the first method
as 3D solution and the second method with fixed $L$ as 2D
solution.

\section{Ground state and metastability in mesoscopic
superconducting disks}

As a typical case let us consider a disk of radius $R$=0.8$\mu
m$, thickness $d$=0.134 $\mu m$, coherence length $\xi(0)$=0.183
$\mu m$ and penetration length $\lambda(0)$=0.07$\mu m$
(resulting in $\kappa=0.383$) which are comparable to those of
one of the $Al$ disks considered in the experiment of Geim {\it
et al} \cite{gei}. In the case of the 2D solution we solve for a
particular $L$ state. The
transition between the different $L$ states is obtained from a
comparison of the free energies from which we obtain the ground
state.  The free energy G as a function of increasing applied
magnetic field is shown by the thin solid curves in Fig.~1(a)
for different $L$ states.  The corresponding magnetization are
shown in Fig.~1(b) by the thin solid curves.  Hence from
Fig.~1(a), it can be seen that up to a magnetic field of 42.6
Gauss, the $L$=0 state is the ground state.  Beyond this field
the $L$=1 state becomes the ground state. As we increase the
field, higher $L$ states become the lowest energy states. This
continues as long as the free energy G is negative. A positive G
means that the normal state has a lower free energy than the
superconducting state.  Hence the free energy of the ground
state of the system is given by the thick solid curve in
Fig.~1(a). The corresponding magnetization along this ground
state is given by the thick solid curve in Fig.~1(b). Notice
that each $L$ state has a metastable region where the sample can
be paramagnetic.

The 3D approach does not have the restriction of fixed $L$ and
consequently many solutions can be found for given magnetic
field which are not necessarily the lowest energy state.  Our
approach is to start at zero magnetic field, find the ground
state, and then increase the magnetic field in small steps in
which we use the order parameter of the previous magnetic field
as the initial state in the iteration of the coupled set of
differential equations. The magnetization as given by the 3D
solution in increasing magnetic field is given by the thick
dotted curve in Fig.~1(b). The corresponding free energy is
given by the thick dotted curve in Fig.~1(a). As compared to the
2D solution we find that (within errors arising due to the
assumption that the order parameter is uniform in the disk in
the z-direction in case of the 3D solution), the 3D solution in
increasing magnetic field, takes the system along a state that
conserves $L$.  This persists up to the point where the $L$=0
state is no longer metastable, i.e. it no longer has a local
minimum. Then the system jumps to the $L$=1 state and starts
evolving along the $L$=1 state until it is no longer stable. 
Following the 2D solution along a path of conserving $L$ we
notice that the line shape of the magnetization and the number
and position of jumps in the magnetization are similar to those
of the 3D solution. The little difference in magnitude arises
due to the fact that the 3D solution uses the extra
approximation of constant order parameter in the z-direction
while the order parameter should be reduced near the upper and
lower surfaces of the disk.  Hence an important conclusion that
can be made from the comparison of the 2D solution and the 3D
solution is that the 2D solution can be used to study the
magnetization of this sample.

Comparing the thick dotted curve in Fig. 1(a) (3D solution) with
the thick solid curve in Fig.  1(a) (the ground state obtained
from 2D solution) we find that the 3D solution in increasing
magnetic field makes the system evolve along metastable states. 
In the 3D solution the metastability is due to energy barriers
which do not allow the sample to switch between different
angular momentum states.  In the Ginzburg-Landau theory there
are two such energy barriers. One of them is a surface barrier
known as the Bean-Livingston barrier  \cite{bea} that has been
well studied in the case of cylinders.  The other barrier is a
volume barrier and it appears only when the sample makes a first
order transition to the normal state.  This appears because
close to the first order transition the free energy has two
minima corresponding to two different values of the order
parameter and there is a barrier separating the two minima. 
These two barriers make the system evolve along metastable
paths.  Therefore, the question arises: what makes the system
evolve along the ground state in increasing magnetic fields? The
volume barriers in these disks are negligibly small for
transitions between different $L$ states of the disk considered
in Fig. 1 because in such a case the order parameter on the
average changes by a small amount or the local minima are very
close to each other.
And as is known for cylinders,
the surface barrier in increasing fields (but not in decreasing
fields) can be destroyed by small surface defects. The disks
used in the experiment of Ref \cite{gei} have a rough boundary
and therefore in increasing field the sample evolves along the
ground state.

The situation is different in decreasing field.  The surface
barrier does not disappear due to surface defects and so in
decreasing field the sample will evolve along metastable states,
thus resulting in hysteresis. 
In our model the barriers are present but what is not
present are the thermal fluctuations in the system.  To a first
approximation we can neglect these effects because of the
extreme low temperature at which the experiments are done and
study hysteresis effects arising due to metastability alone. The
result in decreasing field using our 3D solution is given by
the thick dashed curves in Fig. 1.

\section{Effect of the detector size on the measurement of
magnetization}

The magnetometry used in the experimental work of Ref.
\cite{gei} is explained in detail in Refs. \cite{gei2}. The
superconducting sample is mounted on top of a small Hall cross.
From the Hall resistance one can estimate how much field is
expelled from the Hall cross, which is due to the expulsion of
the field by the superconductor.  The Hall cross has a larger
area than the sample and it measures the magnetization of this
area rather than the sample. In Ref. \cite{lee} it was shown
that the Hall voltage of a Hall bar, in the ballistic regime, is
determined by the average magnetic field piercing through the
Hall cross region.  Due to the depletion of the two dimensional
electron gas near the edges of the leads of the Hall cross the
effective area of the Hall cross is not exactly known and may be
smaller than the optical size.  Since the field distribution in
case of thin disks is extremely non-uniform inside as well as
outside the disks (this is essentially different for cylinders
where the magnetic field outside the sample always equals the
applied field) the detector size will have an effect on the
measured magnitude of the magnetization.  Since the field just
outside the disk is larger than the applied field, a larger
detector will underestimate the magnetization, the nature and
extent will depend on the field profile outside the disk.

To understand these effects we calculate the magnetization of an
area larger than that of the sample by integrating the field
expelled from this area.  We take a disk of radius 0.44 $\mu m$,
thickness = 0.15 $\mu m$, coherence length $\xi(0)$ = 0.275 $\mu
m$ and penetration length $\lambda(0)$ = 0.07 $\mu m$ at
T=0.4K.  In Fig.~2 we plot the calculated flux expulsion from
the sample (solid curve), and from a square area larger (width
1.75 $\mu m$, dashed curve) and smaller (width .067 $\mu m$,
dotted curve) than the sample size.  The detector area is taken
immediately under the sample.  When the detector width is larger
than the diameter of the sample, increase of the detector width
simply scales down the magnetization.  When the detector width
is smaller than the diameter of the sample there is a small
change in the line shape. The dashed curve when multiplied by
15.83 practically coincides with the solid curve, whereas the
dotted curve when scaled by 0.5291 goes over to the dash-dotted
curve.

In Fig. 3 we show how the scale factor (defined as peak value of
magnetization for a certain detector width divided by the peak
value of the sample magnetization) scales with the width $w$ of
the detector. The magnetic field profile (applied field + field
due to the magnetization of the sample) for a disk of radius
$0.44 \mu m$ is shown in the lower inset. From this profile it
is clear that increasing the detector size $w$ will result in an
average magnetic field closer to the applied field and
consequently a smaller measured field expulsion and consequently
a smaller magnetization. The scale factor shown in Fig. 3 could
be fitted to a Gaussian curve with center at 0.18, width of
0.93, height of 1.83 and offset of 0.055. Since the magnetic
field is also non-uniform in the z-direction, the magnetization
measurement will also be affected by the distance of the
detector from the sample.  We take a square detector whose width
is equal to the diameter of the sample and place it at different
distances below the sample. In this case there is no change in
line shape and the different curves can be made to coincide with
each other by appropriate scaling. In the upper inset of Fig. 3
we show how the scale factor varies with the vertical distance
from the middle of the sample. This curve could be fitted to the
function $Aexp[(z-z_0)t]$ where $A$=0.6, $z_0$=0.07 and
$t$=0.28. In section V we will show that for samples with $L>1$
the line shape of the magnetization curve can also change with
the size and position of the detector but the peak value scales
in a similar manner as found in this section.  We shall also
show that in decreasing fields the conclusions can be very
different because magnetization can change sign as a function of
$w$.

\section{``fractional fluxoid disk" and the non-linear Meissner
effect}

If the upper critical field ($H_c$) of a disk is such that $H_c
\pi R^2 < \phi_0$, then we call it a fractional fluxoid disk,
where $\phi_0=hc/2e$ is a single flux quantum.  Physically it
means that the disk is not large enough to accommodate even a
single vortex when in the superconducting state. In such a small
disk the $L$=1 state cannot nucleate and the superconducting
state always corresponds to the $L$=0 state.  In this regime the
2D solution is always valid.  As an example, we fix the radius
of the disk to be 0.3 $\mu m$, temperature T=0.4K, coherence
length $\xi(0)$=0.25 $\mu m$ and penetration length
$\lambda(0)$=0.07 $\mu m$ which results in $\kappa$=0.28. We
plot in Fig. 4 the magnetization versus decreasing applied
magnetic field for 5 different thicknesses $d$.  The values of
$t=d/\xi(T)$ are shown in the figure where $\xi(T)$=0.31 for
$T$=0.4K.  It can be seen that for $t\le$0.3 the disk shows a
second order phase transition to the normal state, whereas for
$t\ge$0.5, we find a first order phase transition to the normal
state.  The magnetization (multiplied by 0.5) from the LGL
theory is shown by the dashed curve for the same radius and
coherence length.  The magnetization for a cylinder in
decreasing magnetic field for the same radius and material
parameters is shown by the dash-dotted curve.  Notice that for
very small thickness $t$ the sample shows a second order
transition between the normal-superconducting state but as the
thickness increases, beyond a critical thickness of t=0.33, this
becomes a first order transition.  The magnitude of the jump of
the magnetization $\Delta M$ versus $t$ at the
superconducting-normal transition point is shown in Fig. 5. This
result can be fitted, (solid curve) to the function
$A|t-t_c|^p$, where $A=2.87\pm0.07$, $t_c=.33$ and
$p=0.56\pm 0.02$. In the inset we plot the peak position of the
magnetization curve versus $t$ which we could fit to the
function $a+bt^c$ where $a=55.40\pm 1.05$, $b=43.91\pm0.99$
and $c=0.59\pm0.03$. It is generally believed that the upper
critical field is given accurately by the LGL theory which we
find to be true in decreasing magnetic fields.

The same results in increasing magnetic field is given in Fig. 6
where curve conventions are the same as in Fig.~4.  In
increasing magnetic field the LGL theory gives the correct upper
critical field as long as the sample shows a second order
transition to the normal state. In this case the upper critical
field is independent of the thickness. As the sample starts
showing a first order transition the upper critical field starts
increasing with increasing thickness. The reason behind this
qualitative difference with the LGL theory in increasing and
decreasing field arises due to the fact that in the LGL theory
there is only one minimum in the free energy whereas in the GL
theory, when the sample shows a second order transition, there
is one minimum but when it shows a first order transition, there
are two minima in the free energy (the first minimum is at zero
order parameter and the second minimum is at a finite order
parameter). One of them is however metastable and we shall soon
find that the experimental system can exist in such a metastable
state.  In this case $\Delta M$ versus $t$ is shown in Fig. 7.
It fits to a function $A(t-t_c)^p$ where $A=2.16$,
$t_c$=0.33 and $p$=0.58.

From Figs. 4 and 6 we can understand why a disk can show a first
order or a second order phase transition to the normal state as
is found in the experiment. It also explains why a fractional
fluxoid disk in case of a first order transition shows a
different behavior in increasing and decreasing magnetic fields
and as a result a remarkable hysteresis as seen in the
experiment.  In Fig.~8, we show that a disk of radius $R$=0.44
$\mu m$, thickness $d$=0.15 $\mu m$, coherence length
$\xi(0)$=0.275 $\mu m$ and penetration length $\lambda(0)$=0.07
$\mu m$ at T=0.4K can exhibit a magnetization like that of a
disk whose magnetization was measured in the experiment and
whose radius was reported as 0.5 $\mu m$, thickness between 0.07
$\mu m$ and 0.15 $\mu m$, coherence length of 0.25 $\mu m$ and
penetration length of 0.07 $\mu m$. Note that we decrease the
coherence length by 10 $\%$ and the radius by 12 $\%$ to
reproduce the experimental result which are well within the
errors of experimental estimates. The increasing field behavior
as seen in the experiment is given by the open squares and the
decreasing field behavior is given by the open circles.  Our
numerical calculation of the magnetization in increasing and
decreasing magnetic field for a detector area of 2.5 $\mu m$
placed 0.15 $\mu m$ below the sample is shown by the solid and
dotted curves, respectively. The inclusion of the detector size
and position results in a magnetization which is a factor 50.44
smaller than the pure sample magnetization. The inset shows the
dimensionless free energy as a function of the magnetic field. 
The dashed line in Fig. 8 is the behavior expected in the case
of a linear Meissner effect and is drawn tangential to the small
field magnetization curve. Notice that even in the intermediate
magnetic field region the magnetization is substantially
non-linear. Note that in decreasing field the position of the
jump does not agree with the experiment. In case of a first
order transition the free energy has two minima separated by a
barrier and when the system tries to switch from one minima to
the other it has to overcome this barrier. The position of the
jump given here corresponds to the situation when this barrier
disappears. But due to fluctuations in the system the jump can be at
lower fields.

The field distribution along a radial line starting from the
center of the disk is shown in Fig.~9(a).  The center of the
disk is at the origin of the coordinates. Notice that the field
is minimum at the center of the disk. It increases drastically
with distance from the center and becomes maximum at 0.44 $\mu
m$ which is precisely the radius of the disk.  This implies that
the field is strongly expelled from the center of the disk. The
11th curve (applied field=70.86 Gauss) corresponds to the
critical field at which superconductivity has disappeared.

Magnetization measurement on bulk samples and large radius
cylinders shows that in the pure ``Meissner state", the
sample behaves as a perfect diamagnet which means magnetization
is proportional to the applied field with a susceptibility
($\chi$=4$\pi M$/H) of -1. This experimental observation led to the
well known London theory. It can already be seen from the
magnetization measurement (see the experimental data in Fig.~8)
that London theory is not valid for the present disks because of
the non-linear magnetization versus the applied field. Even in
the small field region the susceptibility is much smaller than
-1 because of large field penetration in a finite size sample as
can be seen from Fig. 9(a). In the usual London approach the
field penetration is determined by the field value at the
boundary of the sample. In Fig.~9(b) we show the normalized
field distribution, i.e., the field is divided by its maximum
value which is found at the edge of the disk.  This helps us to
compare the field penetration at different applied fields. For
small applied fields (say less than 25 G) the curves fall on top
of each other as expected from London theory.  But for larger
applied fields the field penetration is no longer only
determined by the field at the boundary of the disk and becomes
substantially larger. In the inset of Fig.~9(b) we plot the
magnetic field normalized to the applied field.

\section{Few fluxoid disks}

When a disk is large enough to accommodate one or more fluxoids
we call it a ``few fluxoid disk".  Unlike for ``fractional
fluxoid disk", for the ``few fluxoid disk" case the LGL and GL
theories give qualitatively similar result. The upper critical
field and the number of jumps in the magnetization is the same
in both theories.  But in order to explain the position and the
magnitude of the jumps it is necessary to consider the full GL
eqns.  To explain the increasing field behavior we assume that
the system remains in the ground state, which can be determined
from the 2D solution as explained in section III.

Fig.~10(a) shows the magnetization of a few fluxoid disk: the
open circles are the experimental data (at T=0.4 K) plotted
according to the scale on the left axis.  The experimental
estimates of radius is 0.75 $\mu m$, thickness is between 0.07
$\mu m$ to 0.15 $\mu m$, coherence length is 0.25 $\mu m$ and
penetration length is 0.07 $\mu m$. Sample magnetization along
the ground state calculated from our 2D solution is shown by the
dashed curve and it is plotted according to the scale on the
right.  Parameter values used for the radius is 0.8 $\mu m$,
thickness is 0.134 $\mu m$, temperature is 0.4 K, coherence
length is 0.183 $\mu m$ and penetration length is 0.07 $\mu m$.
There is a factor of 25 difference in the magnitude of
magnetization.  However when we use a detector of width 3 $\mu
m$ placed 0.15 $\mu m$ below the sample, the ground state
magnetization, as defined by the flux expulsion from the
detector square, is given by the solid curve plotted according
to the scale on the left. Notice that the inclusion of the
detector effect explains the magnitude of magnetization and it
also changes slightly the slope of the curve for fields larger
than 40 G. Namely the magnetization decreases faster with
magnetic field.

Also the magnetization in decreasing magnetic field is shown in
Fig.~10(b) where the curve conventions are the same as for the
field sweep up.  It is needless to say that the agreement with
the experimental data is remarkable except for the last jump at
low field.  In Fig. 11 we show how the minimum magnetization at
55.795 Gauss scales with the size of the detector in case of the
field down sweep. For detector size smaller than the sample the
magnetization is strongly paramagnetic due to the pinned fluxoid
in the sample. Also when the detector size slightly exceeds the
sample size the magnetization can be paramagnetic which decrease
as the detector size is made larger.

Next in Fig. 12 we consider an even larger disk which is able to
contain 19 flux quanta before it becomes normal. The
experimental data of magnetization versus increasing magnetic
fields at T=0.4 K is shown by the open circles according to the
scale on the left. The experimental estimate of the parameters
are radius=1.2 $\mu m$, thickness is between 0.07 $\mu m$ and
0.15 $\mu m$.  Our 2D solution for the ground state
magnetization of the sample is shown by the dashed curve
according to the scale on the right. Parameter values used for
radius is 1.25 $\mu m$, thickness is 0.136 $\mu m$, coherence
length is 0.195 $\mu m$ and penetration length is 0.07 $\mu m$. 
When we use a detector of width 4 $\mu m$ placed 0.15 $\mu m$
below the sample the ground state magnetization is given by the
solid curve. Note that the size of the detector used here
appears to be a bit larger than expected experimentally. Using a
detector of smaller width does not appreciably change the line
shape as long as it is slightly larger than the disk diameter.
Also the convex shape of the curve for field larger than 25G is
recovered although it is less than that found experimentally.
Note that the position of the first jump, the upper critical
field and the total number of jumps is the same as that found
experimentally.

The jumps observed in Figs.~10 and 12 reminds us of the Little
Parks oscillations seen in ring shaped superconductors and other
multiply connected geometries \cite{mos}. In such geometries it
occurs due to flux quantization in superconductors and hence
these jumps occur at extremely regular intervals (provided the
sample is clean).  It is to be noted that in Figs.~10 and 12 the
interval between jumps is slowly decreasing in the experimental
data as well as in the theoretical curve.  In Fig.~13(a) we plot
the magnetic field interval between two successive jumps,
$\Delta H$, versus the angular momentum $L$ for the ground state
(open circles) which are compared with the experimental data
(solid squares). We find that the experimental $\Delta H$ is
extremely irregular, which suggests that entry of fluxoids into
the sample is hindered by disorder and pinning centers.  Another
reason may be the experimental accuracy and fluctuations in the
applied field which are expected to lead to an error of about 10
$\%$ in these quantities.  The overall behavior: of a large
$\Delta H$ for small $L$ and a smaller but constant $\Delta H$
for larger $L$ is clearly visible in the experimental result. 
The open squares in Fig. 13(a) are obtained from the recent
theoretical result of Ref. \cite{pal}.  He used a variational
approach to solve the GL equation with the approximation of an
uniform magnetic field distribution. His calculation was done
for $R=5.25\xi$ and $\kappa = 1.2$.  The origin of this drastic
decrease in $\Delta H$ arises due to strong flux expulsion and
the demagnetization effects at low fields because of which the
flux inside the sample increases at a very slow rate as compared
to the applied field. So the external field has to be increased
by more than a flux quantum to give rise to one flux quantum
inside the disk.  Such strong flux expulsion becomes weaker as
field is increased and becomes zero at the critical field. So at
higher fields $\Delta M$ versus $L$ becomes flat and corresponds 
to the flux quantization condition: $\Delta H \pi R^2=\phi_0$.  

In Fig.~13(b) the magnitude of the jumps in the magnetization
($\Delta M$) is shown and compared with the experimental data
and the theoretical result of Ref. \cite{pal}. The theory gives
a substantial larger magnitude of the jumps than seen
experimentally except for the smaller radius sample (see inset)
when $L>4$.

\section{Conclusions}

Apart from explaining quantitatively and qualitatively the
recent experiments of Geim {\it et al} we can make some general
conclusions.  The GL theory captures all the features of
mesoscopic superconductors provided the non-linear term and the
field distribution is properly accounted for. Hence it is
necessary to review the calculation of various quantities in
mesoscopic superconductors made so far which did not account for
these effects.  We can also conclude that in increasing magnetic
fields the sample evolves along the free energy minimum and the
Bean-Livingston barrier plays no role. Surface defects destroy
the Bean-Livingston barrier in increasing fields in the case of
cylinders. This also seems to be true for disks. We also find
that the magnitude of magnetization is drastically affected by
the detector size. The nature of the measured magnetization curves are
sensitive to small changes in parameters.

\section{Acknowledgments}

This work is supported by the Flemish Science Foundation(FWO-Vl)
grant No: G.0232.96, the Belgian Inter-University Attraction
Poles (IUAP-VI).  One of us (PSD) is supported by a post
doctoral scholarship from the FWO-Vl and FMP is a Research
Director with the FWO-Vl.  It is a pleasure to acknowledge
stimulating discussions with A. Geim.

\centerline{FIGURE CAPTIONS}

\noindent FIG. 1. Free energy versus increasing magnetic field
for different angular momentum ($L$) states are shown by thin
solid curves.  Thick solid curve is the free energy versus
increasing magnetic field for the ground state of the sample,
the thick dotted curve is the result
from the 3D solution and the thick dashed curve
shows the same for decreasing magnetic fields.  Here $G_0$=
$H_c^2V/8 \pi$. (b) The corresponding magnetization for the
curves in (a) with the same convention for the different curves.

\noindent FIG.~2. Magnetization versus increasing magnetic field
for different detector sizes.

\noindent FIG. 3. The scale factor for the magnetization as
function of the width of the detector. The upper inset shows how
this scale factor varies with the distance of the detector.  The
lower inset shows how the magnetic field varies along a line
from the center of the disk for the particular disk considered
in this figure.

\noindent FIG.~4. The gradual change from a first order to a
second order phase transition as the thickness of a sample is
increased. The curves are obtained for decreasing magnetic field
sweep.

\noindent FIG. 5. $\Delta M$ is the magnitude of the jumps in
the magnetization seen in Fig. 4. The solid circles give $\Delta
M$ for the corresponding value of $t$. The solid curve is a
$\chi^2$-fit to it. The inset shows how the peak position of the
curves in Fig. 4 vary with $t$ along with a $\chi^2$-fit. 

\noindent FIG.~6. The gradual change from a first order to a
second order phase transition as the thickness of a sample is
increased for increasing magnetic field sweep.

\noindent FIG. 7. The solid circles give $\Delta M$ for the
corresponding value of $t$. The solid curve is $\chi^2$ fit to
it. The inset shows how the transition field vary with $t$ along
with a $\chi^2$ fit.

\noindent FIG.~8. Comparison between experimentally observed
magnetization in increasing (open squares) and decreasing fields
(open circles) and our numerically calculated magnetization in
increasing (solid curve) and decreasing (dotted curve) magnetic
field.  The dashed curve is a tangent to the experimental data
at the origin. The inset shows the dimensionless free energy for
the system.

\noindent FIG. 9. (a) Magnetic field change from the center of
the disk considered in Fig. 8 for different applied fields
mentioned in the figure. (b) Magnetic field curves of Fig. 9(a)
when scaled by the maximum field for each curve. The inset
shows the magnetic field when scaled by the
applied field.

\noindent FIG.~10. (a) Comparison between experimentally
observed magnetization in increasing (open circles) fields
according to the scale on the left y-axis and our numerically
calculated magnetization in increasing (dashed curve) magnetic
field according to the scale on the right  y-axis. The solid
curve is according to the left y-axis and is obtained when the
detector size is taken into account. (b) The same for decreasing
magnetic fields.

\noindent FIG. 11. The magnetization of the sample of Fig. 10(b)
at  55.795 Gauss as function of the detector size.

\noindent FIG.~12. Comparison between experimentally observed
magnetization in increasing (open circles) fields according to
the scale on the left y-axis and our numerically calculated
magnetization in increasing (dashed curve) magnetic field
according to the scale on the right  y-axis. The solid curve is
according to the left y-axis and is obtained when the detector
size is taken into account.

\noindent FIG.~13. (a) Comparison of experimentally observed and
numerically obtained $\Delta H$ (magnetic field change due to penetration
of one extra fluxoid) versus $L$
for the disk considered in Fig. 12. The
solid squares give the experimental data, the 
open circles are our numerically obtained values and the
open squares correspond to the results
of Ref. \cite{pal}. The curves are guides to the eye.
The inset shows the same for the case of
Fig. 10(a). (b) Comparison of experimentally observed and
numerically obtained $\Delta M$ (jumps in the magnetization) versus $L$
for the disk considered in Fig. 12. The symbol conventions are the same as in
Fig. 13(a).
The inset shows the same for the case of
Fig. 10(a).

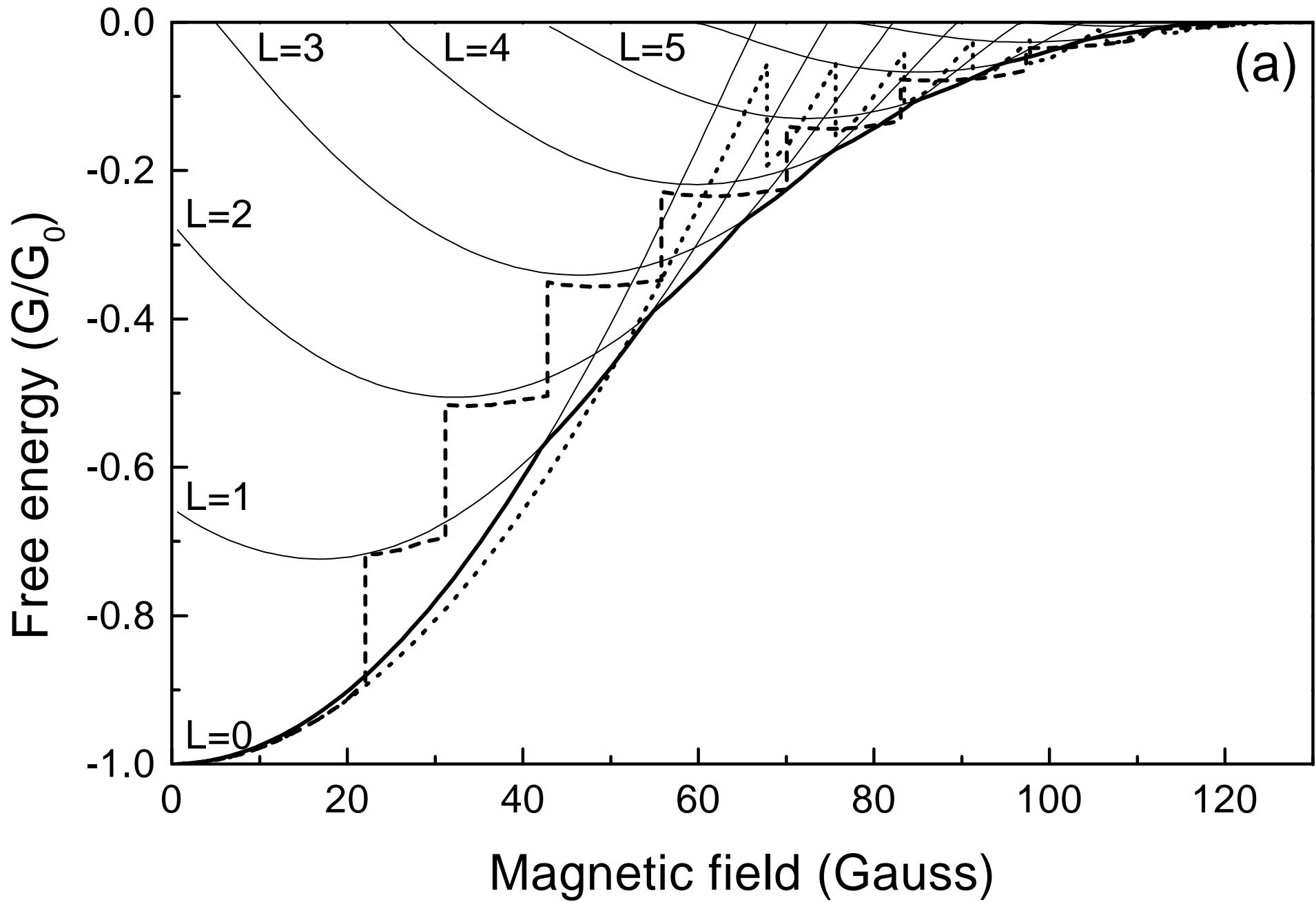
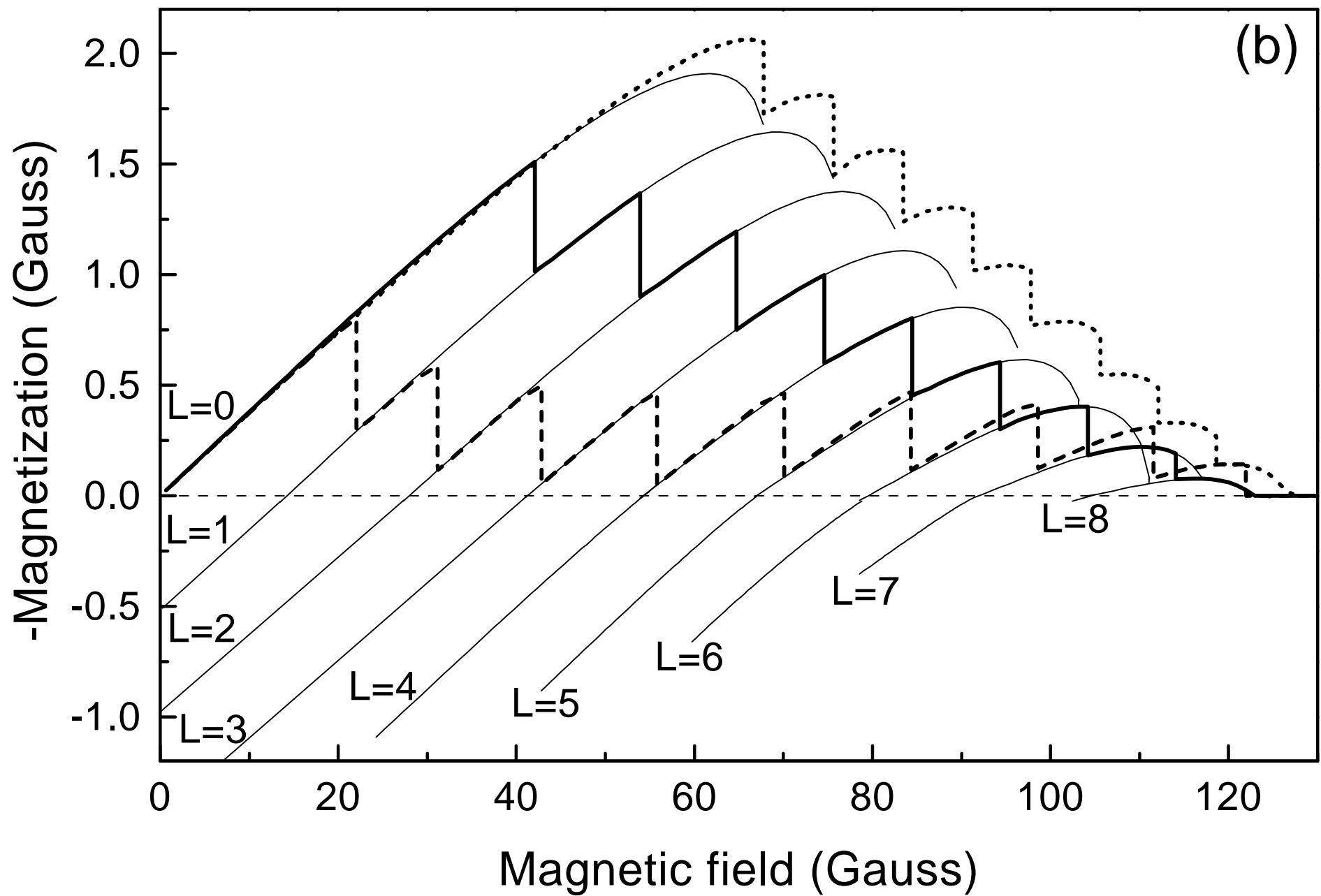
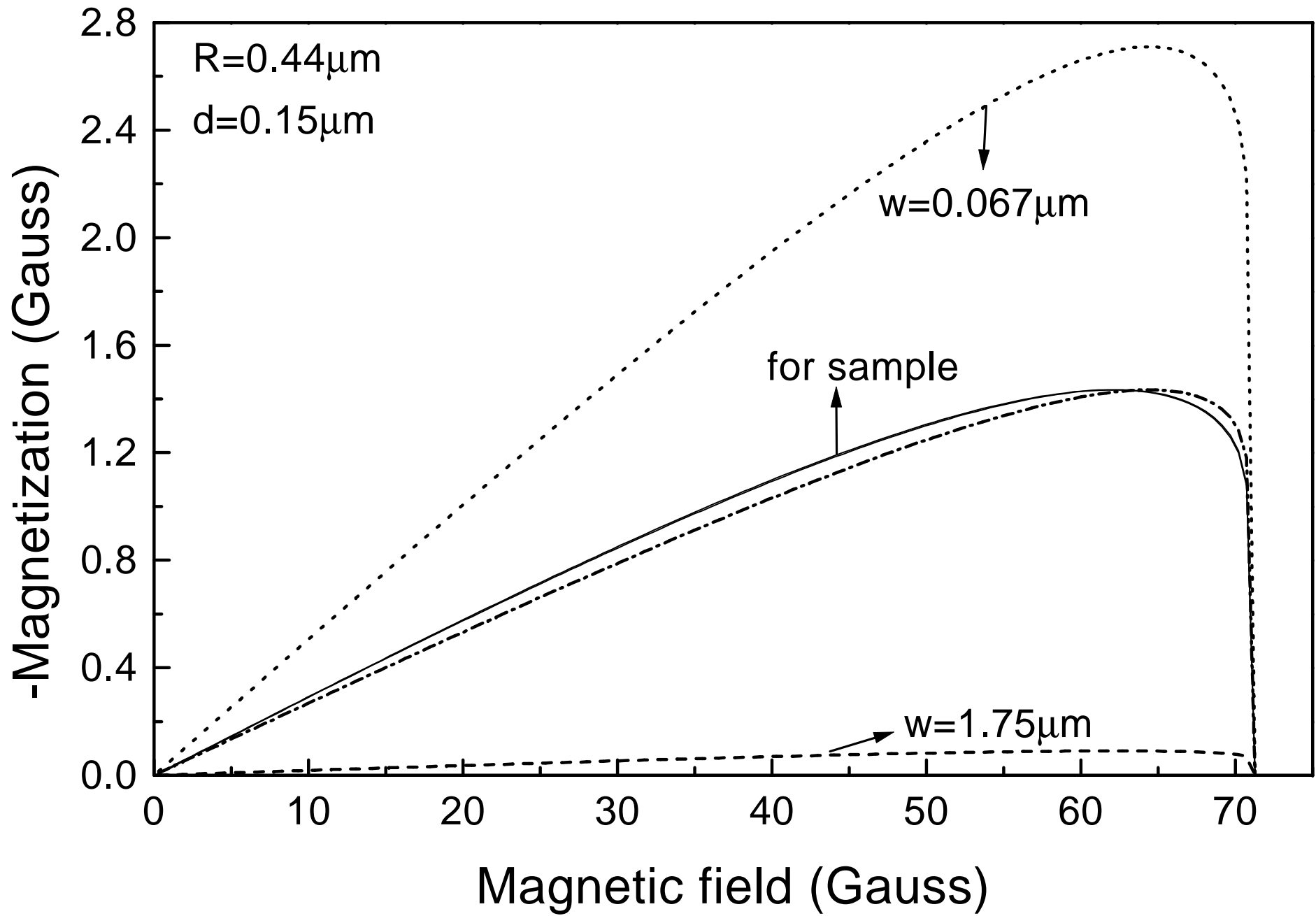
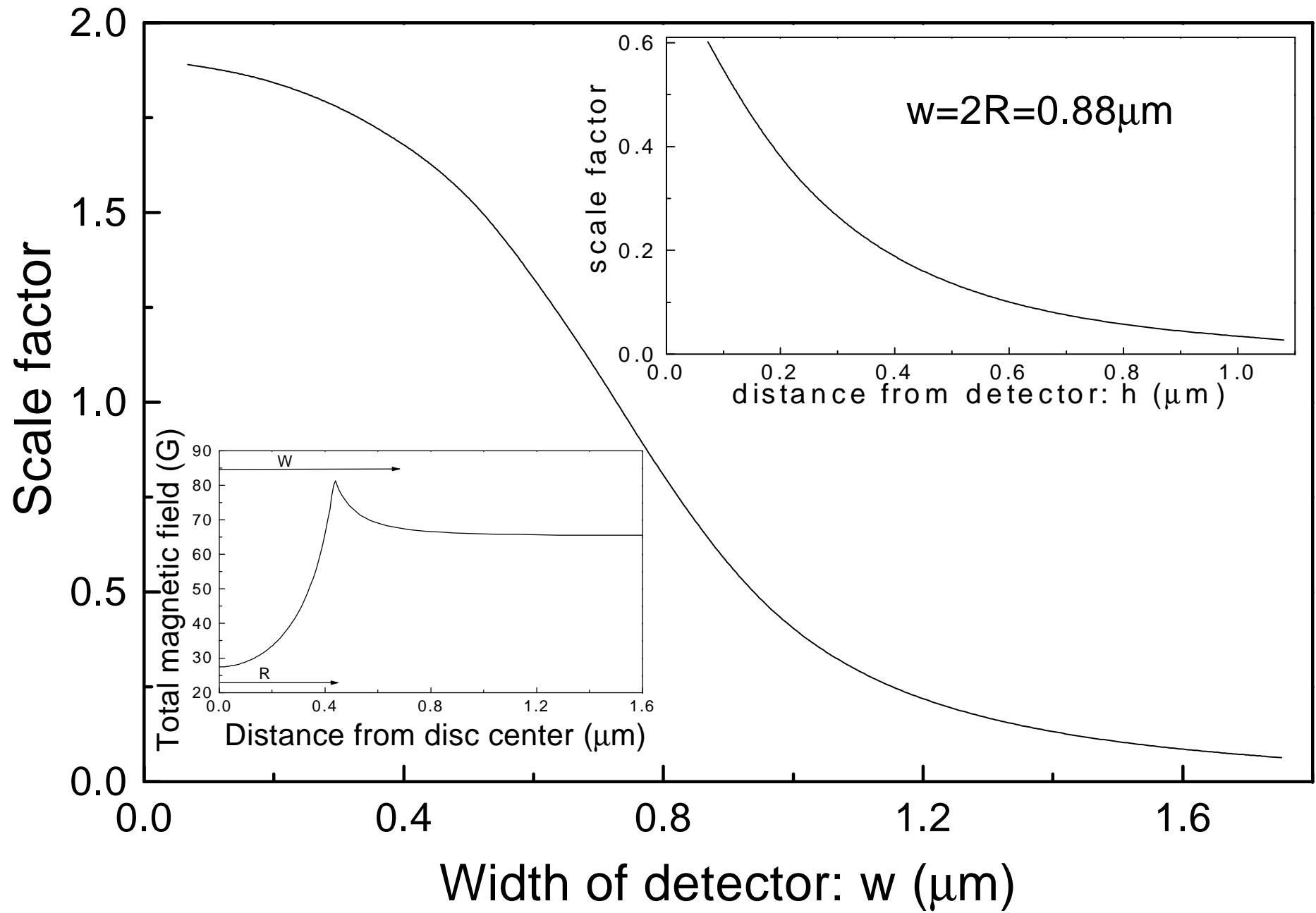
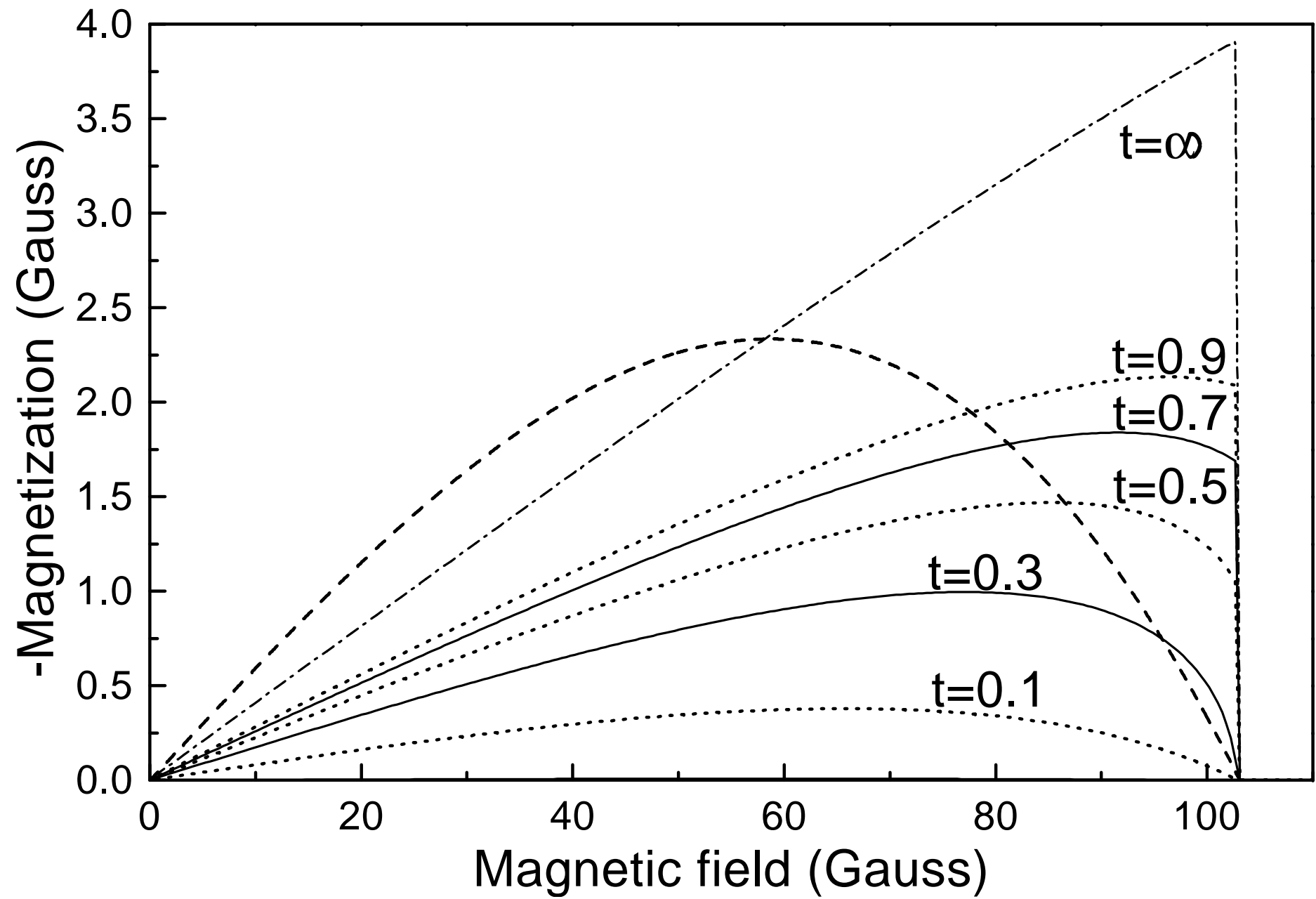
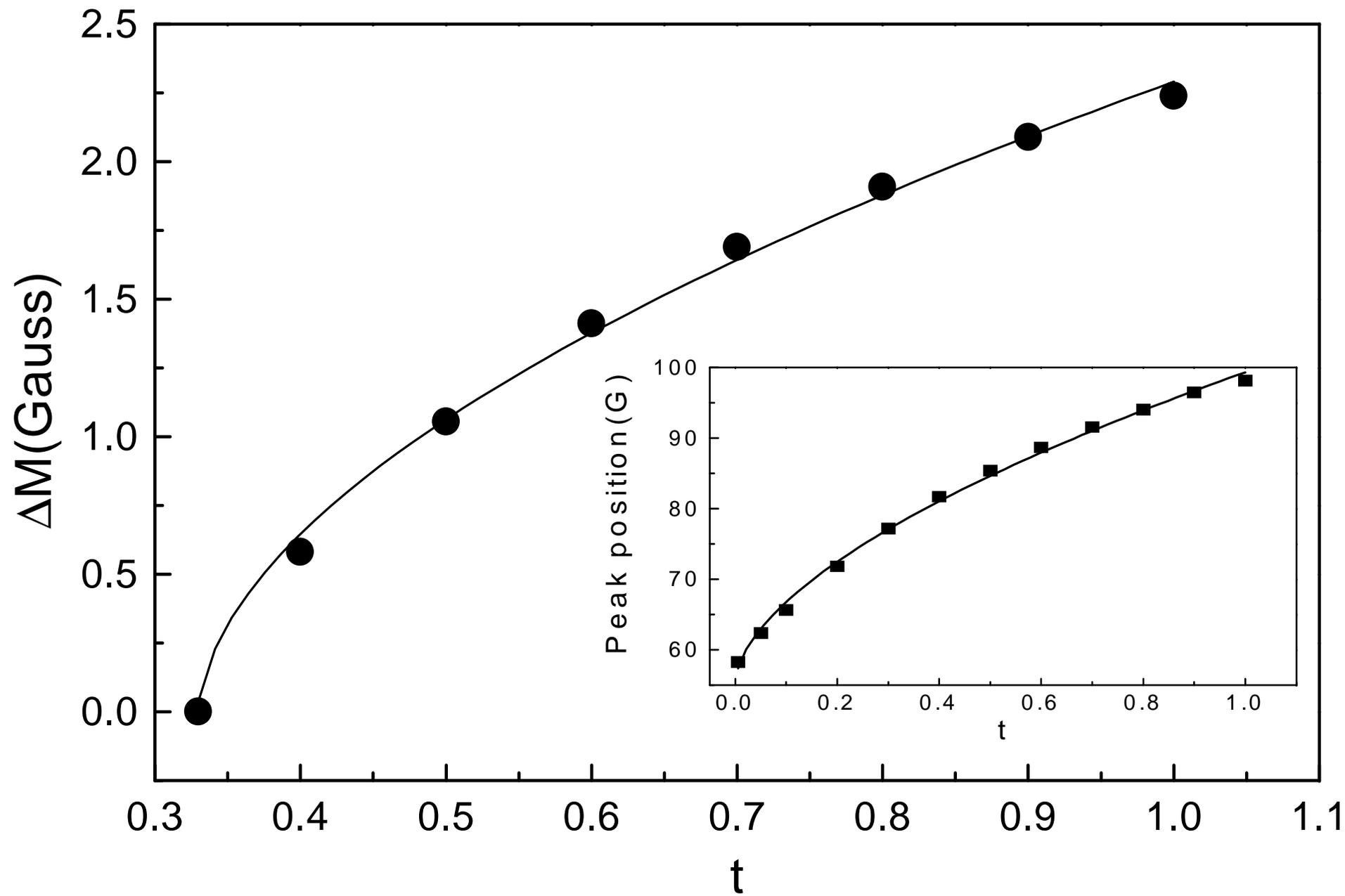
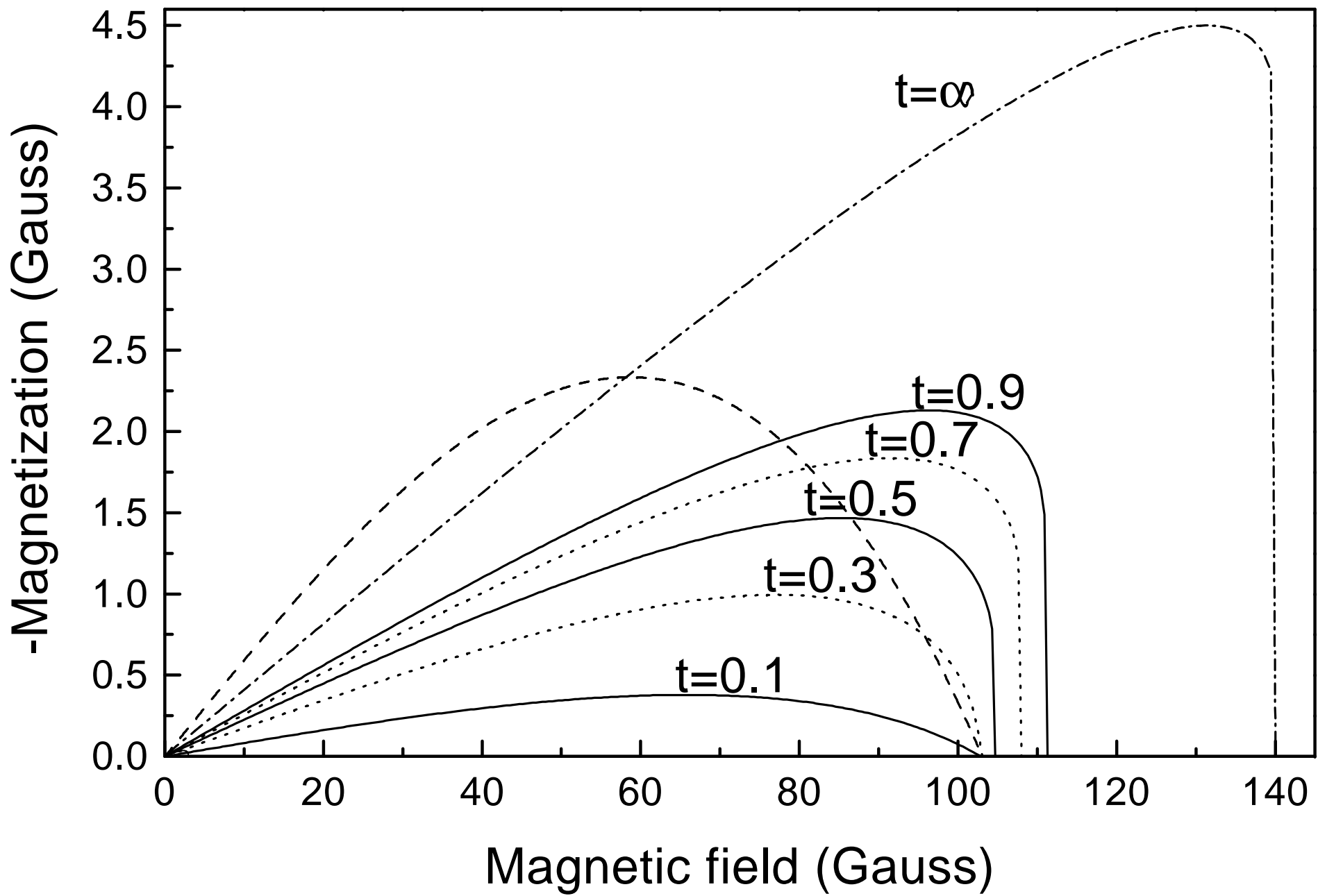
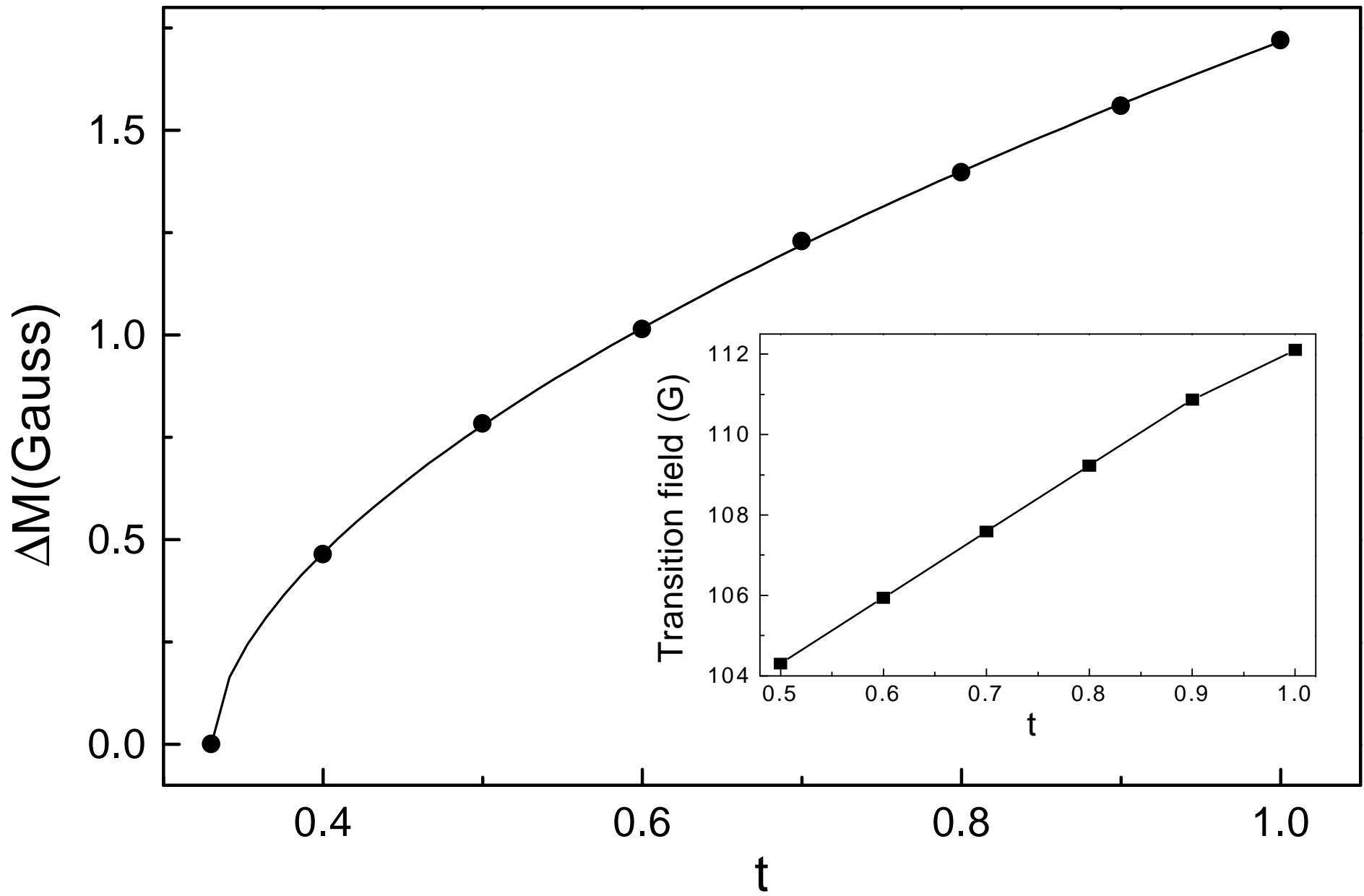
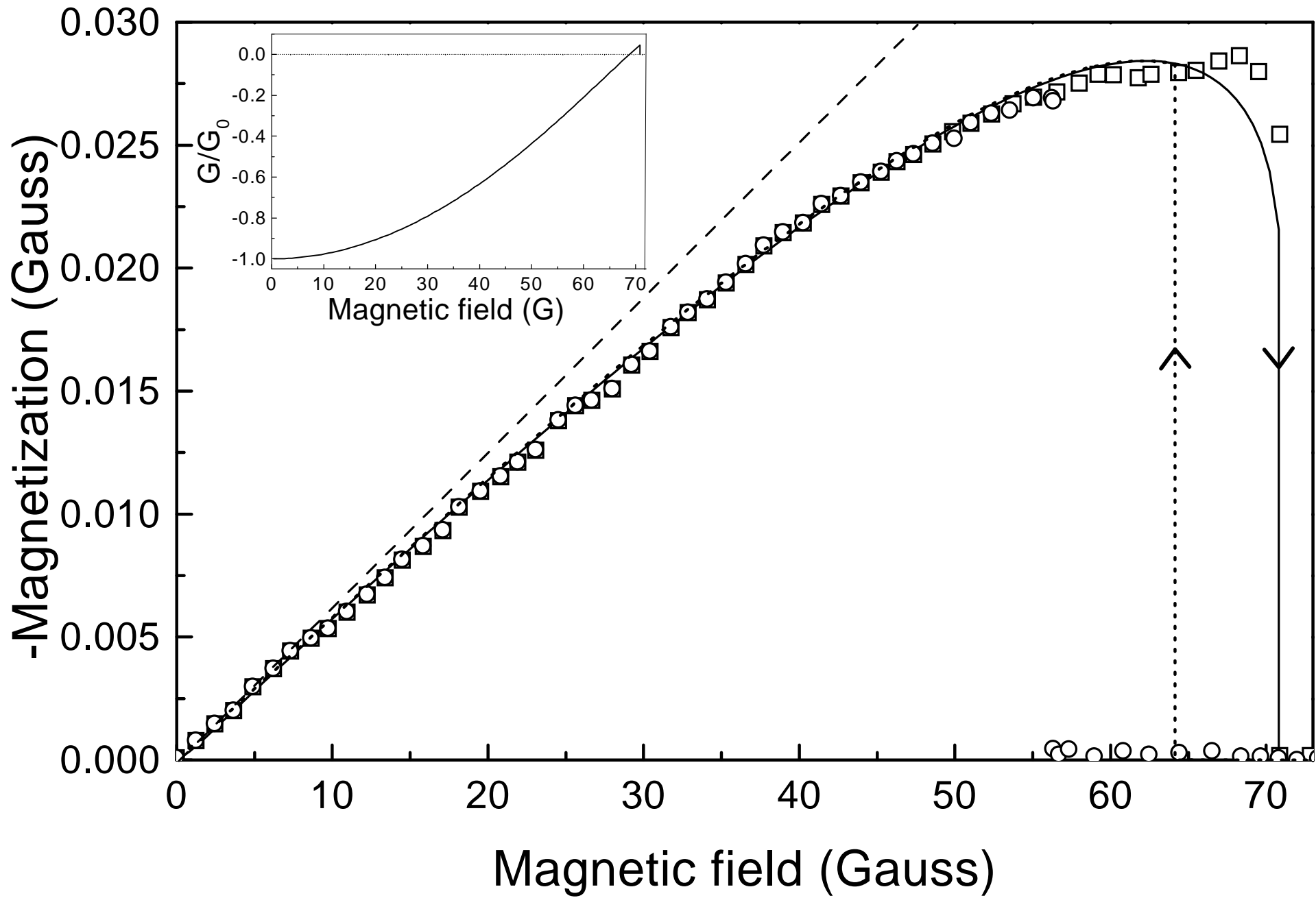
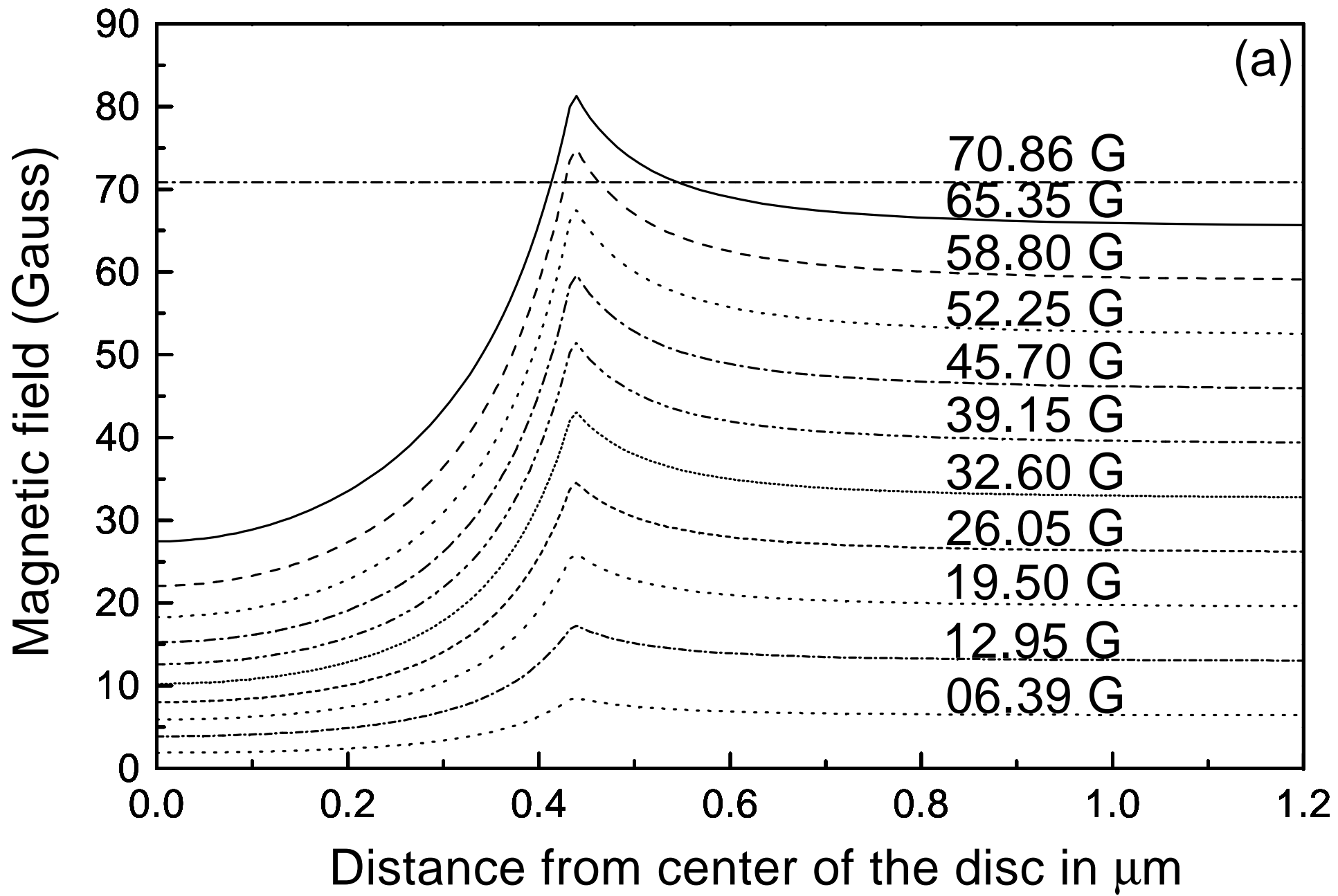
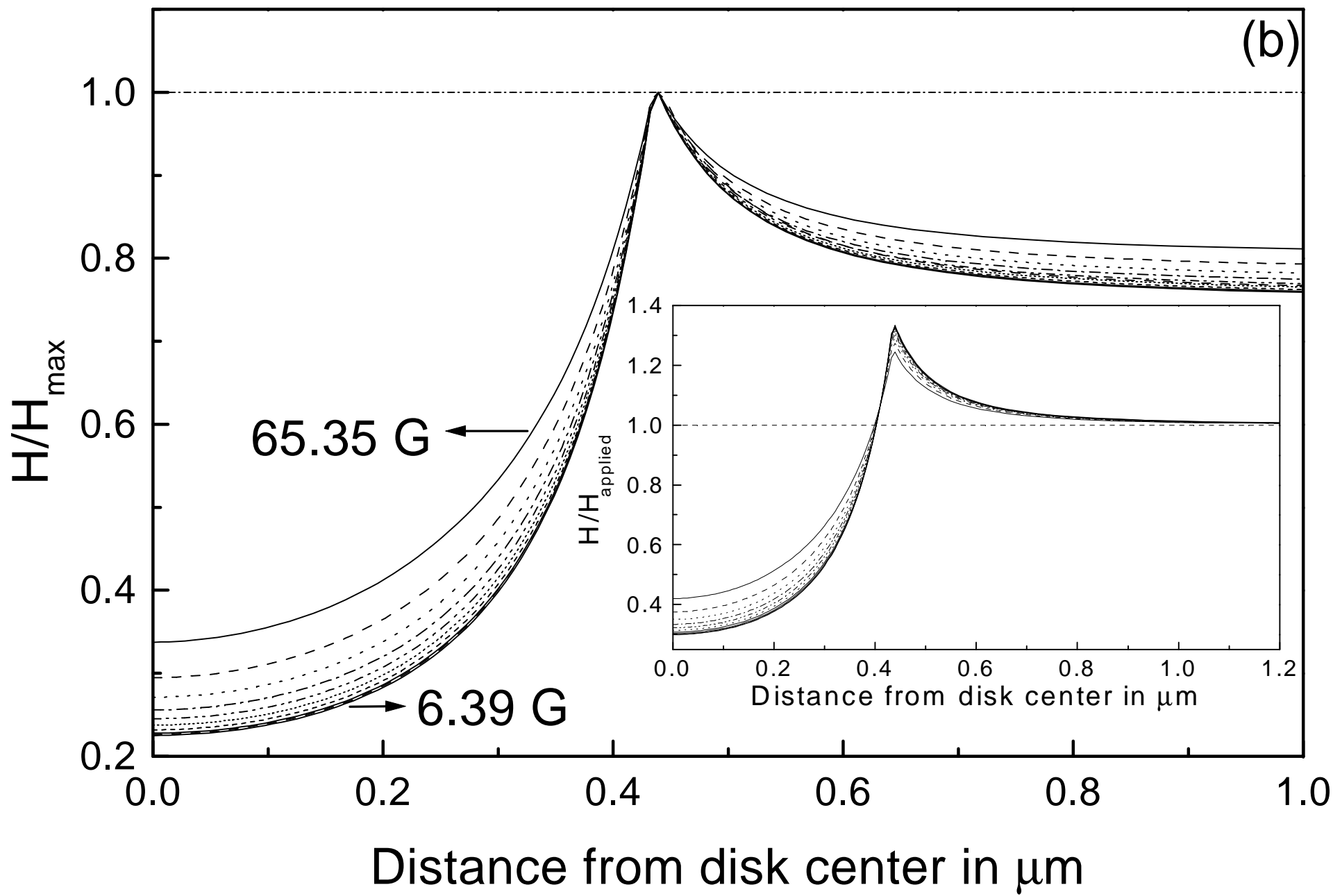
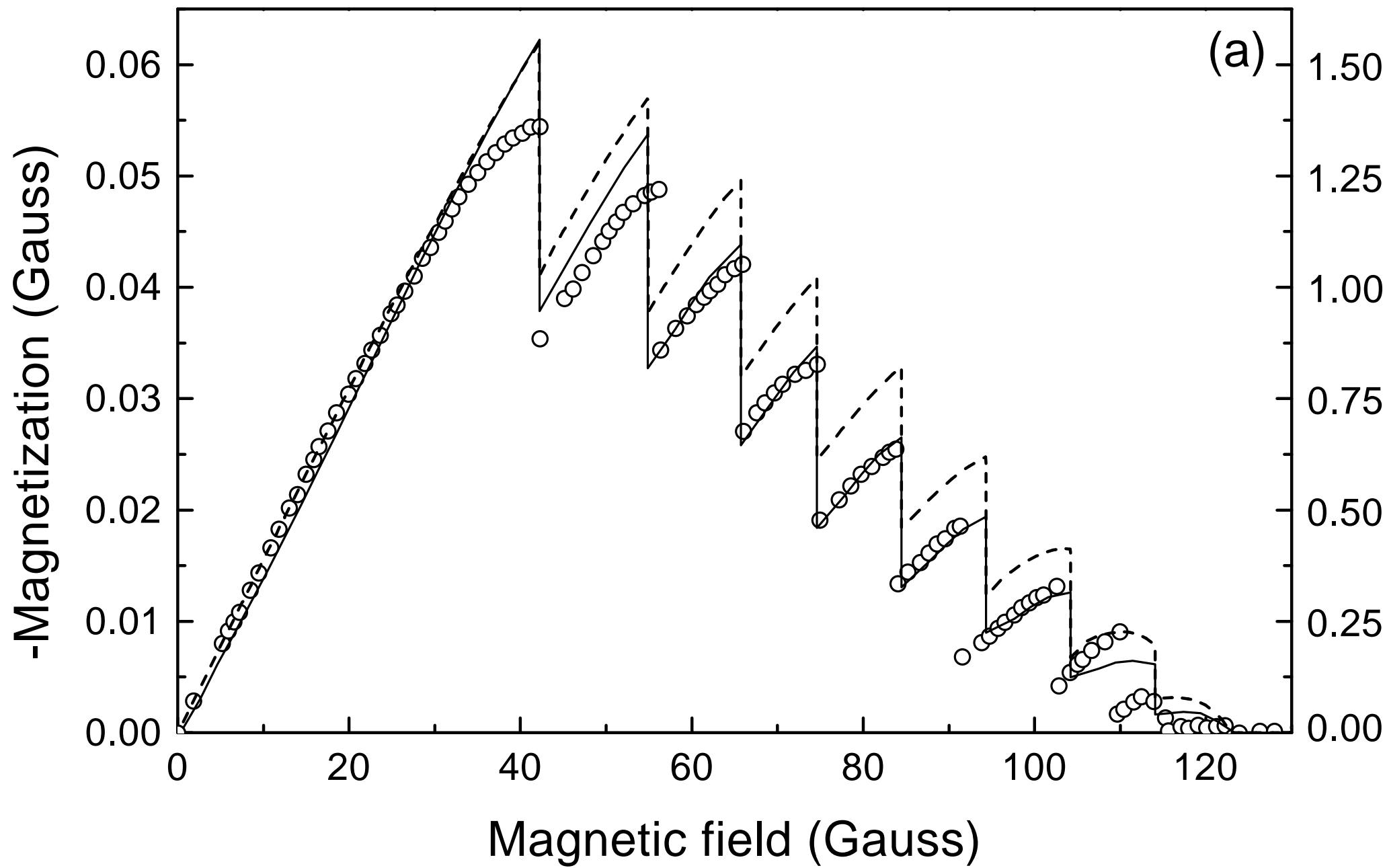
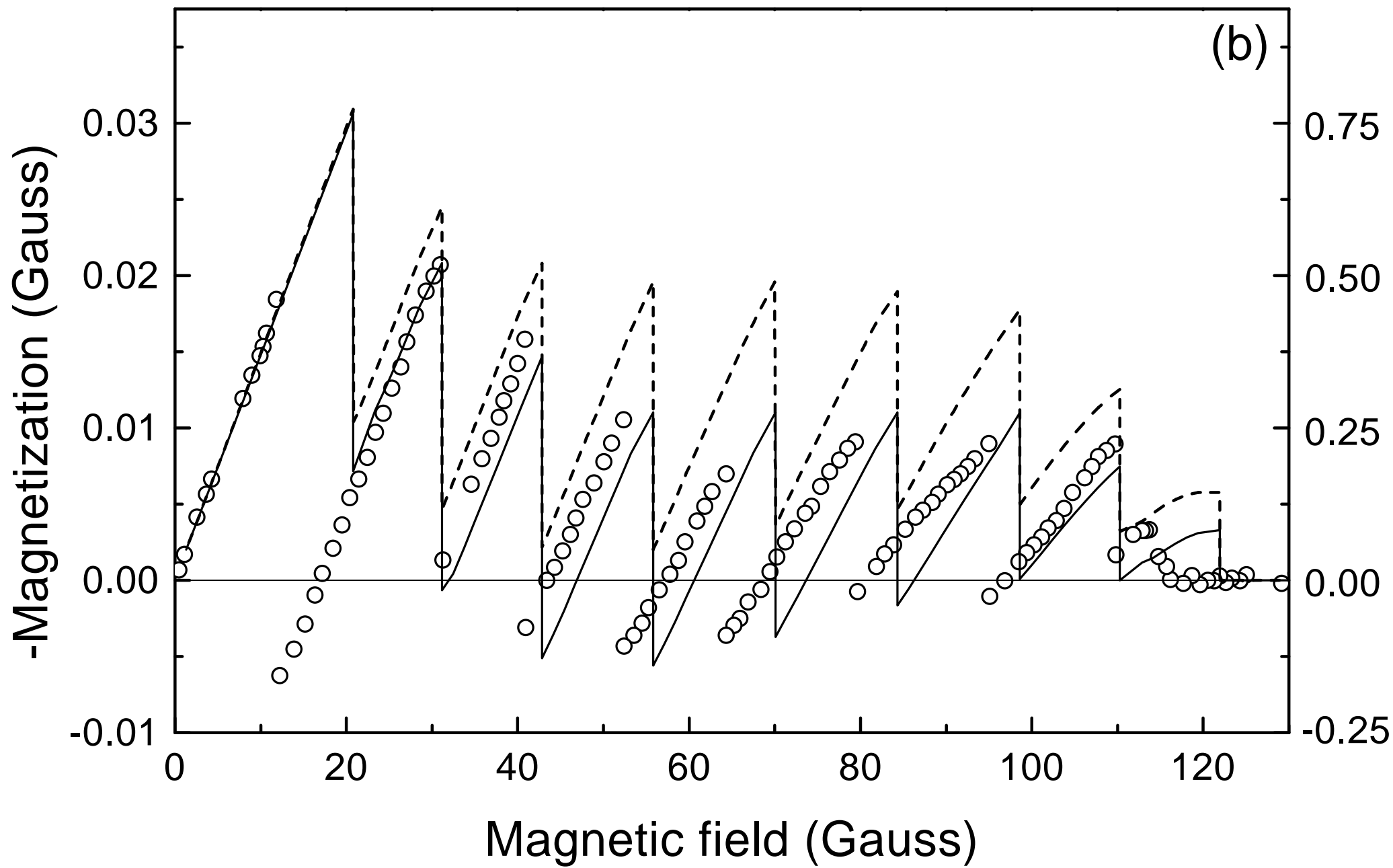
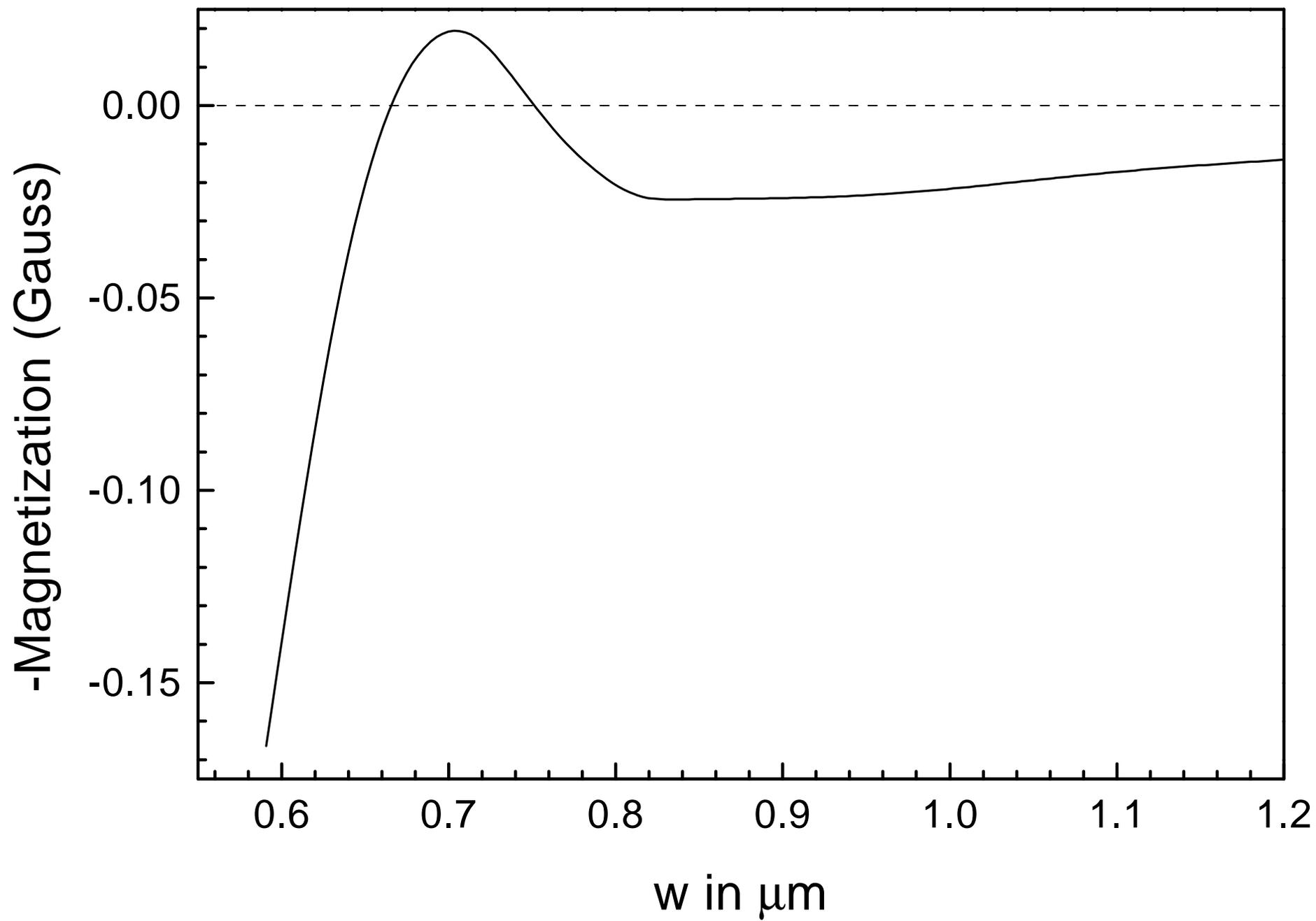
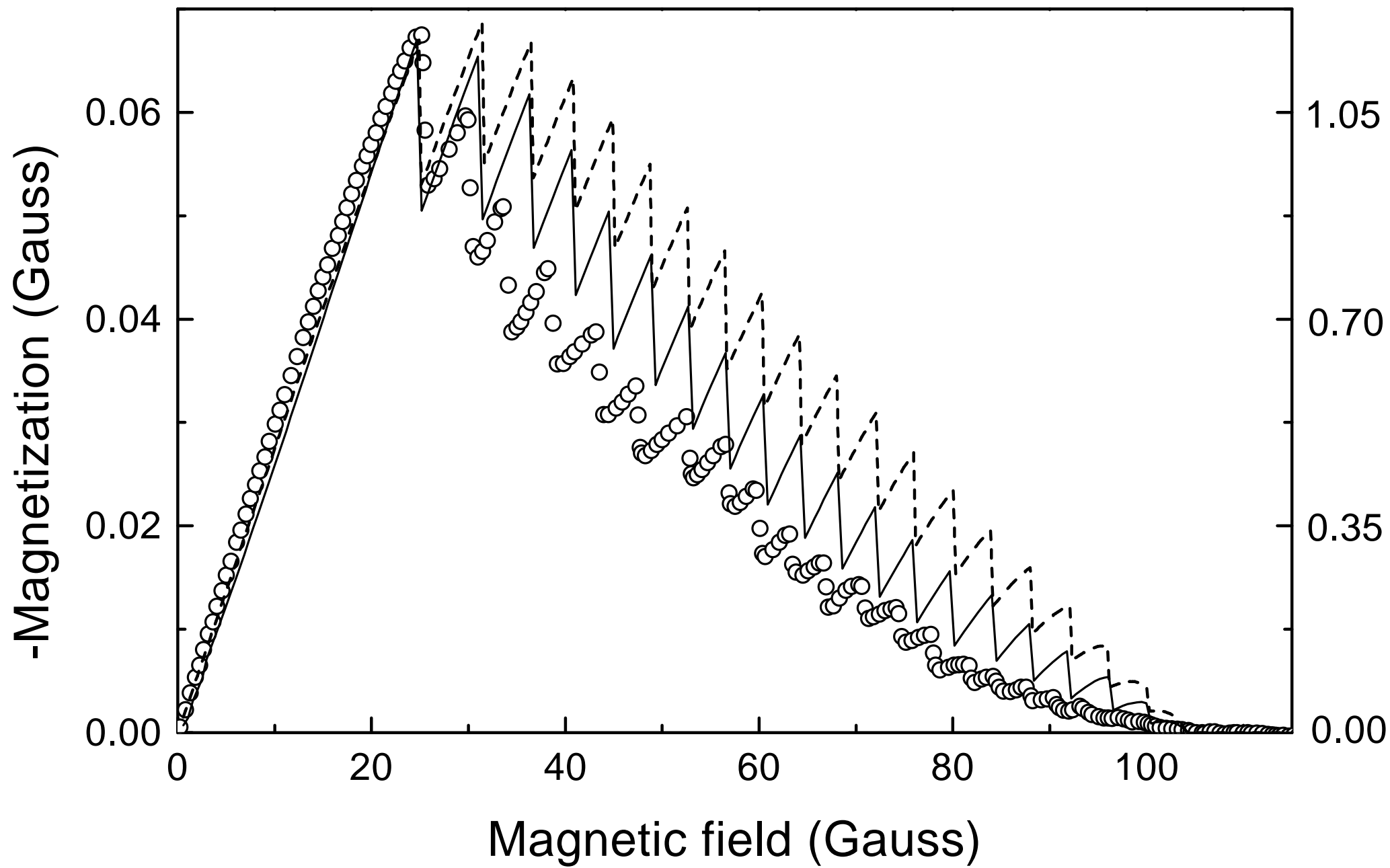
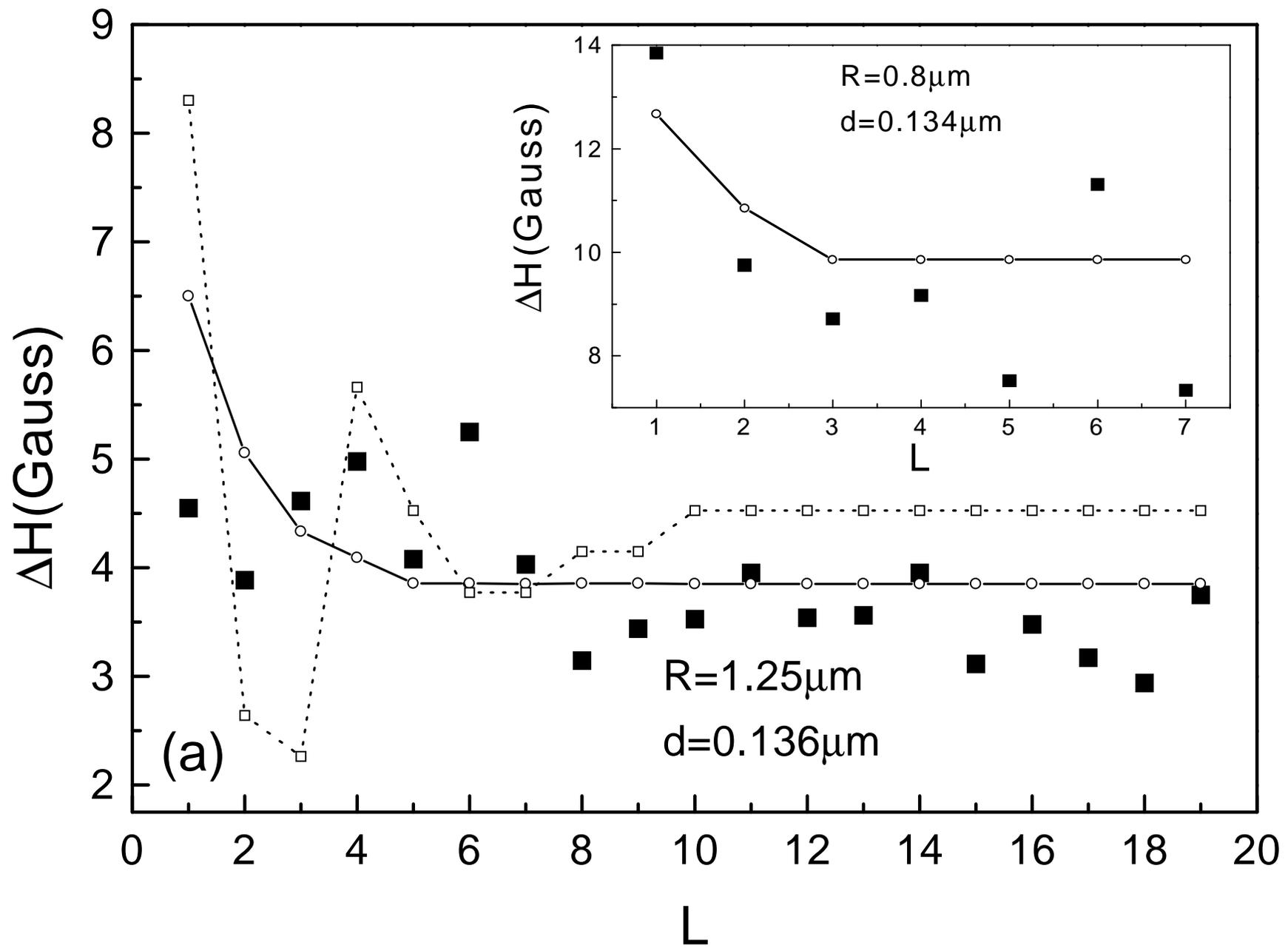
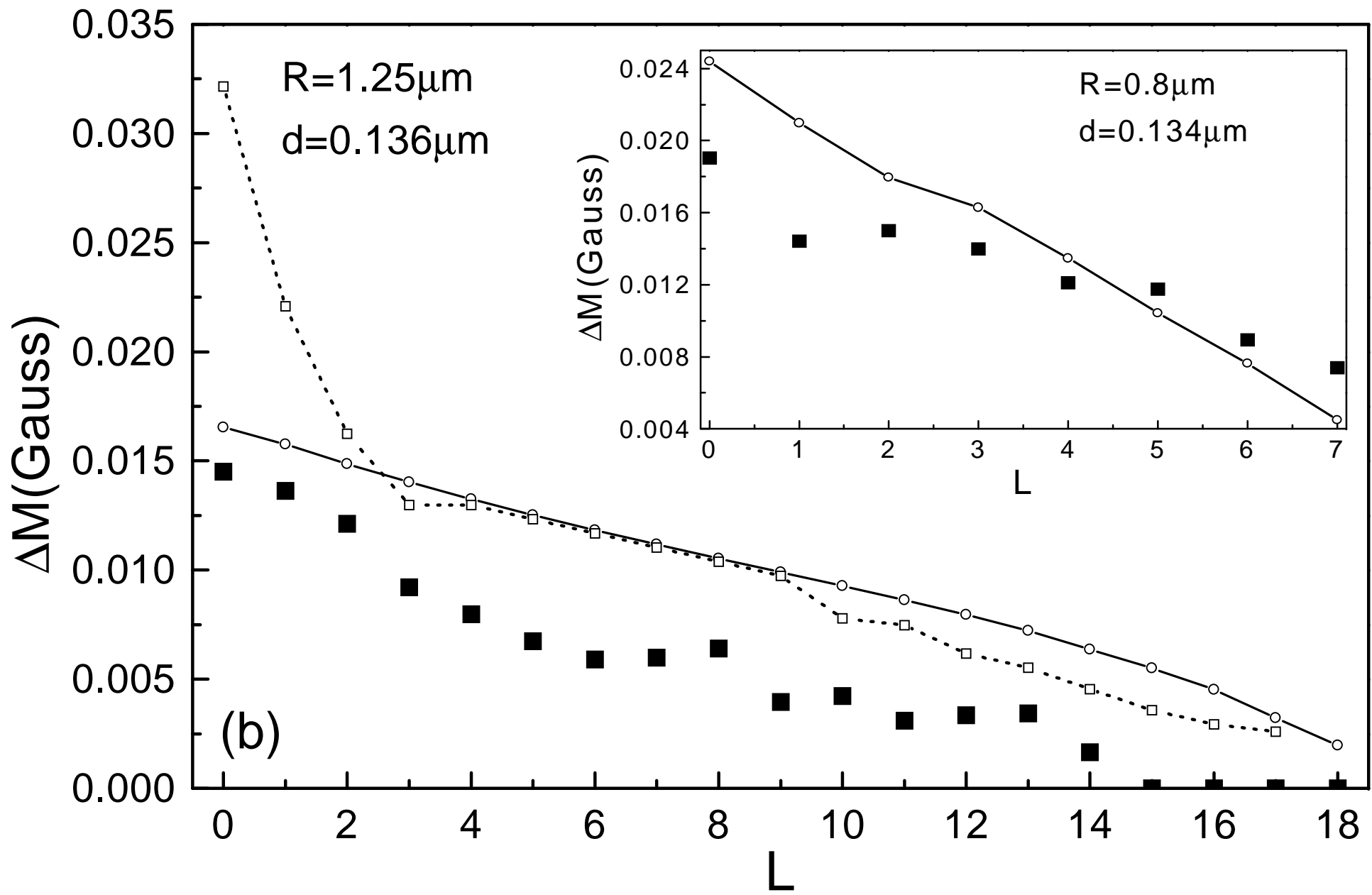
\end{document}